\documentclass[a4paper]{article}
\usepackage{spconf,amsmath,graphicx,cite}

\usepackage{amsfonts, float, bm, url}

\floatstyle{ruled} \newfloat{algorithm}{tbp}{loa}
\providecommand{\algorithmname}{Algorithm}
\floatname{algorithm}{\protect\algorithmname}

\usepackage{algpseudocode}

\newcommand{\R}{\mathbb{R}}

\newcommand{\setX}{\mathcal{X}}

\newcommand{\ui}{[0,1)} 
\newcommand{\uid}{\ui^d}

\newcommand{\E}{\mathbb{E}}
\newcommand{\var}{\mathrm{Var}}

\newcommand{\Unif}{\mathcal{U}} 

\newcommand{\bx}{\mathbf{x}}
\newcommand{\bu}{\mathbf{u}}
\newcommand{\bv}{\mathbf{v}}
\newcommand{\by}{\mathbf{y}}

\newcommand{\dd}{\mathrm{d}}
\newcommand{\dx}{\dd \mathbf{x}}
\newcommand{\du}{\dd \mathbf{u}}
\newcommand{\bigO}{\mathcal{O}} 
\newcommand{\smallo}{{\scriptscriptstyle\mathcal{O}}} 

\newcommand{\Dst}{D^\star} 

\newcommand{\comment}[1]{ \ifthenelse{ \equal{\showcomment}{true} }{ {\bf #1} }{} }
\newcommand{\showcomment}{true}

\newcommand{\IHSFC}{h} 

\title{Application of Sequential Quasi-Monte Carlo to autonomous positioning}
\twoauthors
  {Nicolas Chopin \sthanks{The first author is partially supported by  grant
ANR-11-IDEX-0003/Labex Ecodec/ANR-11-LABX-0047 from
the French National Research Agency (ANR) as part of the ``Investissements d'Avenir''
program.}}
	{CREST-ENSAE\\
	92\,245 Malakoff\\
	France}
  {Mathieu Gerber 
\sthanks{The second author is supported by DARPA under Grant No. FA8750-14-2-0117.}}
	{Harvard University\\
	Department of Statistics\\
	Cambridge, MA 
	}

\begin{document}

\maketitle
\begin{abstract}
  Sequential Monte Carlo algorithms (also known as particle filters) are popular
  methods to approximate filtering (and related) distributions of state-space
  models. However, they converge at the slow $1/\sqrt{N}$ rate, which may be an issue
  in real-time data-intensive scenarios.  We give a brief outline of SQMC (Sequential
  Quasi-Monte Carlo), a variant of SMC based on low-discrepancy point sets proposed
  by \cite{SQMC}, which converges at a faster rate, and we illustrate the greater
  performance of SQMC on autonomous positioning problems.
\end{abstract}
\section{Introduction}

Many problems in signal processing (and related fields) can be formalised as the
filtering of data $(\by_t)$ to recover an unobserved signal $(\bx_t)$ that follows a
state-space model. For non-linear and/or non-Gaussian state-space models, particle
filtering \cite{DouFreiGor,CapMouRyd}, also known as Sequential Monte Carlo (SMC), is now the
standard approach to perform filtering; see e.g. \cite{Djuric2004}.  However, a
potential drawback of SMC for real time applications is its slow $1/\sqrt{N}$
convergence rate (based on $N$ simulations, or `particles').  In real time problems,
the running time per iteration of the filtering algorithm is bounded by the time
interval between successive observations and, consequently, this slow convergence
rate implies that in some settings the approximation error of SMC might be non
negligible.

Recently, \cite{SQMC} proposed and studied the sequential quasi-Monte Carlo (SQMC)
algorithm, which is a quasi-Monte Carlo (QMC) version of particle filtering. Based on
$N$ particles, SQMC has the advantage to converge at rate $\smallo(1/\sqrt{N})$,
i.e. at a faster rate than SMC; see Theorem 7 of \cite{SQMC}. On the other hand, SQMC
requires $\bigO(N\log N)$ operations and is thus slower than SMC, which has
complexity $\bigO(N)$. But \cite{SQMC} show that, in several scenarios, the faster
convergence of SQMC does more than compensate its slower running time and,
consequently, for a given computational budget, SQMC typically achieves a
significantly smaller error size than SMC.

In this paper we propose to apply SQMC to the problem of autonomous positioning of a
vehicle moving along a two dimensional space where, following \cite{Miguez2004}, we
assume that the Markov transition is non Gaussian.  Our numerical study show that for
this real time application SQMC provides a much more accurate estimation of the
position of the vehicle than SMC.

\section{Sequential quasi-Monte Carlo}

\subsection{Background on sequential Monte Carlo}

To introduce SMC we consider the following generic state-space model, described
in term of probability density functions:
\begin{equation}\label{eq:GenericSS}
\begin{cases} 
\by_t|\bx_t\sim f^Y(\by_t|\bx_t),&t\geq 0\\ \bx_t|\bx_{t-1}\sim
f^X(\bx_t|\bx_{t-1}),&t\geq 1\\ \bx_0\sim f_0^X(\bx_0)
\end{cases}
\end{equation} 
where $(\bx_t)_{t\geq 0}$ is the unobservable Markov process on
$\setX\subseteq\mathbb{R}^d$ and $(\by_t)_{t\geq 0}$ is the observation process.

The typical quantity of interest in state-space models is the filtering
distribution, that is, the distribution of $\bx_t$ given all the available
observations at time $t$, which is given by
\begin{equation}\label{eq:filtering}
\begin{split} &p(\bx_t|\by_{0:t}) = \frac{1}{Z_t} \times 
\\ &
\int_{\setX^t}
f_0^X(\bx_0) \prod_{s=1}^{t} f^X(\bx_s|\bx_{s-1}) 
\prod_{s=0}^t f^Y(\by_s|\bx_s)
\dx_{0:t-1} 
\end{split}
\end{equation} 
where $Z_{t}$ is a normalising constant. Except in linear
Gaussian models, the integrals in \eqref{eq:filtering} are not tractable, 
but one may instead run a particle filter to sequentially approximate
$p(\bx_t|\by_{0:t})$.

The basic idea of particle filtering is to use the Markov transition
$f^X(\bx_{t}|\bx_{t-1})$ to propagate the discrete approximation (for $t\geq 1$)
\begin{multline*}
p^N(\bx_{t-1}|\by_{0:t-1})= \sum_{n=1}^N
W_{t-1}^n\delta_{\bx_{t-1}^n}(\dx_{t-1}), \\
\mbox{with} \sum_{n=1}^NW_{t-1}^n=1,\, W^n_{t-1}\geq 0
\end{multline*}
of the filtering distribution at time $t-1$ to the approximation
\begin{align}\label{eq:Pred}
p^N(\bx_{t-1:t}|\by_{0:t-1})=p^N(\bx_{t-1}|\by_{0:t-1})f^X(\bx_{t}|\bx_{t-1})
\end{align} 
of $p(\bx_{t-1:t}|\by_{0:t-1})$. 
Then, the marginal distribution of
$\bx_{t}$ with respect to
\begin{align}\label{eq:filtering2} 
\tilde{p}^N(\bx_{t-1:t}|\by_{0:t})\propto
p^N(\bx_{t-1:t}|\by_{0:t-1})f^Y(\by_{t}|\bx_{t})
\end{align} 
may be used as an approximation of the filtering distribution at time
$t$. Thus, one can perform an importance sampling step, with proposal
distribution \eqref{eq:Pred} and target distribution \eqref{eq:filtering2}, to
get a weighted particle system $\{W_{t}^{n},\bx_{t}^n\}_{n=1}^N$ which is
approximately distributed from $p(\bx_{t}|\by_{0:t})$; see Algorithm
\ref{alg:SMC} for a more precise description of particle filtering.

\begin{algorithm}[H]
\caption{SMC Algorithm (Boostrap filter)\label{alg:SMC}}
\begin{algorithmic} 
\State Operations must be performed for all $n\in 1:N$
\State Sample $\bx_{0}^{n}$ from $f_0^X(\bx_{0})$ and compute
$W_{0}^{n}=f^Y(\by_{0}|\bx_{0}^{n})/\sum_{m=1}^{N}f^Y(\by_0|\bx_{0}^{m})$
\For{$t=1,\dots, T$} 
\State\label{step:Resampling} 
Sample $a_{t-1}^n$ from $\mathcal{M}(W^{1:N}_{t-1})$, 
the multinomial distribution that produces outcome
$m$ with probability $W_{t-1}^m$ \State Sample $\bx_{t}^{n}$ from
$f^X(\bx_t|\bx_{t-1}^{a_{t-1}^n})$ and compute
$W_{t}^{n}=f^Y(\by_{t}|\bx_{t}^{n})/\sum_{m=1}^{N}f^Y(\by_t|\bx_{t}^{m})$ 
\EndFor
\end{algorithmic}
\end{algorithm}

\subsection{Background on quasi-Monte Carlo}

Loosely speaking, a QMC point set $\bu^{1:N}$ in $\ui^d$ is a set of (deterministic) points
which are ``more uniformly'' distributed than uniform random variates. The most
classical measure of uniformity in the QMC literature is the so called star
discrepancy, defined by
\[
\Dst(\bu^{1:N})=\sup_{\bm{b}\in(0,1)^d}\left|\frac{1}{N}\sum_{n=1}^{N}\mathbb{I}\left(\bu^{n}\in[\bm{0},\bm{b}]\right)-\prod_{i=1}^{d}b_{i}\right|,
\] 
where $\bm{b}=(b_1,\ldots,b_d)$. 
We say that $\bu^{1:N}$ is a QMC point set if
$\Dst(\bu^{1:N})=\bigO(N^{-1}(\log N)^d)$.

The main motivation for using low discrepancy point sets in numerical
integration is the Koksma\textendash{}Hlawka inequality:
\[
\left|\frac{1}{N}\sum_{n=1}^{N}\varphi(\bu^{n})-\int_{[0,1)^{d}}\varphi(\bu)\,\du\right|\leq
V(\varphi)\Dst(\bu^{1:N})
\] 
which explicitly links the integration error and the equidistribution
property of the point set at hand, because the quantity $V(\varphi)$ only
depends on the integrand $\varphi$; see e.g. Chap. 5 of \cite{Lemieux:MCandQMCSampling}
for a definition of $V(\varphi)$. 

A useful variant to QMC is randomised QMC (RQMC), which combines the advantages
of random sampling and of QMC strategies. A RQMC point set $\bu^{1:N}$
is such that $\bu^n\sim\Unif(\ui^d)$ for all $n\in 1:N$ and
$\Dst(\bu^{1:N})=\bigO(N^{-1}(\log N)^d)$ with probability one.  A particularly
interesting construction of RQMC point sets is the nested scrambled method for
$(t,m,s)$-nets (see e.g. \cite{dick2010digital}, Chap. 4, for a definition)
proposed by \cite{Owen1995}, which allows to approximate the integral of smooth
functions with an error of size $\bigO(N^{-1.5+\epsilon})$ for any $\epsilon>0$
\cite{Owen1997b}. In addition, and contrary to plain QMC, no smoothness
assumptions on the integrand $\varphi$ are needed for scrambled net quadrature
rules to outperform Monte Carlo integration \cite{Owen1997b}. This last point is
particularly important in the context of SMC because the resampling step (Step
\ref{step:Resampling} of Algorithm \ref{alg:SMC}) introduces discontinuities
which can not be efficiently handled by deterministic QMC strategies.

\subsection{Sequential quasi-Monte Carlo}

The basic idea of SQMC is to replace the sampling step from the proposal
distribution \eqref{eq:Pred} by a low discrepancy point set with respect to the
same distribution.

The classical way to transform a low discrepancy point set with respect to the
uniform distribution (i.e. a QMC point set) into a low discrepancy point set with
respect to a non-uniform distribution $\pi(\bx)$ on $\setX\subset\mathbb{R}^d$ is to
use the inverse of the Rosenblatt transformation of $\pi$, defined by
\[
F_{\pi}(\bx)=\left(u_{1},\ldots,u_{d}\right)^T,\quad\bx=(x_{1},\ldots,x_{d})^T\in\setX,
\] where, $u_{1}=F_{\pi,1}(x_{1}),$ $F_{\pi,1}$ being the CDF of the marginal
distribution of the first component (relative to $\pi$), and for $i\geq2$,
$u_{i}=F_{\pi,i}(x_{i}|x_{1:i-1})$, $F_{\pi,i}(\cdot|x_{1:i-1})$ being the CDF
of component $x_{i}$, conditional on $(x_{1},\ldots,x_{i-1}$), relative to
$\pi$.

Following this idea, and assuming for the moment that the state variable $\bx_t$ is
univariate, one can generate a low discrepancy point set
$(\hat{\bx}_{t-1}^{1:N},\bx_{t}^{1:N})$ from \eqref{eq:Pred} as follows: let
$\bu_t^{1:N}$ be a (R)QMC point set in $\ui^2$, with $\bu^n=(u_t^n,v_t^n)$, and
compute
$$
\hat{\bx}_{t-1}^n=F^{-1}_{p^N(\bx_{t-1}|\by_{0:t})}(u_t^n),\quad
\bx_{t}^n=F^{-1}_{f^X(\cdot{ }|\hat{x}_{t-1}^n)}(\bm{v}_t^n).
$$
However, the extension of this approach to $d>1$ is not trivial because
the distribution $p^N(\bx_t|\by_{0:t})\dx_t$ is then a (weighted) sum of Dirac
measures over $\R^d$. 

\begin{figure} \centering
\label{fig:hilbert}
\includegraphics[scale=0.7]{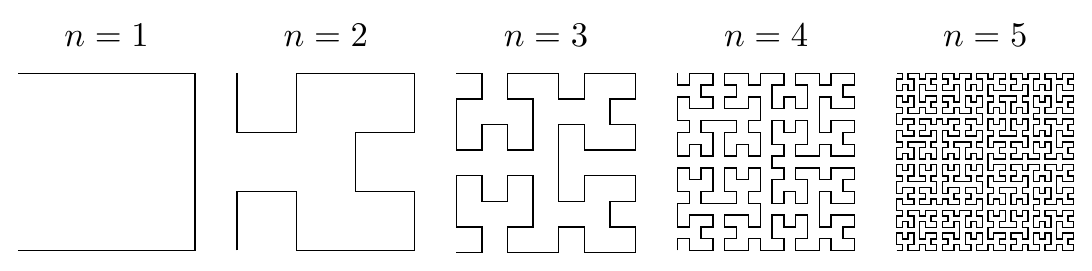}
\caption{The Hilbert curve is a $[0,1]\rightarrow[0,1]^d$ continuous fractal map,
which is obtained as the limit of sequence $(H_n)$, the first elements of which
are represented above (for $d=2$). Source: Marc van Dongen}
\end{figure}

To overcome this difficulty, \cite{SQMC} proposes to transform the multivariate
(discrete) distribution $p^N(\bx_{t-1}|\by_{0:t-1})\dx_t$ into a univariate
(discrete) distribution $p_h^N(h_t|\by_{0:t})\dd h_t$ on $\ui$ using the following
change of variable
$$
\bx\in\setX\mapsto h\circ \psi_t(\bx)\in [0,1],
$$
where $h:[0,1]^d\rightarrow [0,1]$ is the (generalised) inverse of the Hilbert space
filling curve $H:[0,1]\rightarrow [0,1]^d$, 
and $\psi_t:\setX\rightarrow[0,1]^{d}$ is
some user-chosen discrepancy-preserving bijection between $\setX$ and
$\psi_t(\setX)\subset[0,1]^{d}$. See \cite{SQMC} and Section
\ref{sec:simulation-set-up} below for more details on how to choose $\psi_t$,
and see Figure \ref{fig:hilbert} for a depiction of the Hilbert curve in two dimensions.

Using this change of variable, we can see iteration $t$ of SMC as an
importance sampling step form
\begin{multline}\label{eq:Predh}
p_h^N(h_{t-1},\bx_{t}|\by_{0:t-1}) = \\
p_h^N(h_{t-1}|\by_{0:t-1})f^X\left(\bx_{t}|H(h_{t-1})\right)
\end{multline} 
to
$$
p_h^N(h_{t-1}, \bx_{t}|\by_{0:t})\propto
p_h^N(h_{t-1},\bx_{t}|\by_{0:t-1})f^Y(\by_{t}|\bx_{t})
$$
and we can therefore generate a low discrepancy point set
$(\hat{h}_{t-1}^{1:N},\bx_{t}^{1:N})$ from \eqref{eq:Predh} as follows: let
$\bu_t^{1:N}$ be a (R)QMC point set in $\ui^{d+1}$, with
$\bu^n=(u_t^n,\bm{v}_t^n)$, and compute
$$
\hat{h}_{t-1}=F^{-1}_{p_h^N(\cdot{}|\by_{0:t-1})}(u_t^n),\quad
\hat{\bx}_{t-1}^n=H(h_{t-1}^n),
$$
$$
\bx_{t}^n=F^{-1}_{f^X(\cdot{}|\hat{\bx}_{t-1}^n)}(\bm{v}_t^n).
$$
See Algorithm \ref{alg:SQMC} for a pseudo-code description of SQMC.

\begin{algorithm}\label{alg:SQMC}
\caption{SQMC Algorithm (Boostrap filter)\label{alg:SQMC}}
\begin{algorithmic}[1] 
\State Operations must be performed for all $n\in 1:N$
\State Generate a QMC point set $\bu_{0}^{1:N}$ in $\ui^{d}$ 
\State Compute
$\bx_{0}^{n}=F_{f_0^X}^{-1}(\bu_{0}^{n})$ and
$W_{0}^{n}=f^Y(\by_{0}|\bx_{0}^{n})/\sum_{m=1}^{N}f^Y(\by_0|\bx_{0}^{m})$

\For{$t=1,\dots, T$} 
\State Generate a QMC point set $\bu_{t}^{1:N}$ in
$\ui^{d+1}$, let
$\bu_{t}^{n}=(u_{t}^{n},\bv^n_t)$, with $u_t^n\in\ui$, $\bv_t^n\in\uid$
\State Find permutation $\tau$ such that
$u_t^{\tau(1)} \leq\ldots\leq u_t^{\tau(N)}$ 
\State\label{step1} Hilbert sort:
find permutation $\sigma_{t-1}$ such that
$$
\IHSFC\circ\psi_{t-1}(\bx_{t-1}^{\sigma_{t-1}(1)})
\leq\ldots\leq\IHSFC\circ\psi_{t-1}(\bx_{t-1}^{\sigma_{t-1}(N)})
$$
\State Compute $a_{t-1}^n=F_{t,N}^{-1}(u_t^{\tau(n)})$ where
$F_{t,N}(m)=\sum_{n=1}^NW^{\sigma_{t-1}(n)}_{t-1}\mathbb{I}(n\leq m)$ 
\State Compute\label{step2}
$\bx_{t}^{n}=F_{f^X(\cdot|\bx_{t-1}^{a_{t-1}^n})}^{-1}(\bv_{t}^{\tau(n)})$ and

$W_{t}^{n}=f^Y(\by_{t}|\bx_{t}^{n})/\sum_{m=1}^{N}f^Y(\by_t|\bx_{t}^{m})$ 
\EndFor
\end{algorithmic}
\end{algorithm}

\subsection{Practical implementation}

The complexity of Algorithm \ref{alg:SQMC} is $\bigO(N\log N)$, because it performs
two sorting steps at each iteration.  Regarding the practical implementation of
Algorithm \ref{alg:SQMC}, note that: (a) QMC generation (Steps 2 and 5) routines are
available in most software (e.g. package \texttt{randtoolbox} in \texttt{R}, or class
\texttt{qrandset} in the Statistics toolbox of \texttt{Matlab}); to compute the
$a_{t-1}^n$'s (Step 8), one may use the standard approach based on sorted uniforms
for resampling; and (c) in order to compute $h$, see e.g. \cite{Hamilton2008b}, and
Chris Hamilton's C++ program available at
\url{https://web.cs.dal.ca/~chamilto/hilbert/index.html}.

Our SQMC implementation is available at
\url{https://bitbucket.org/mgerber/sqmc}. We shall use RQMC (randomised QMC) point
sets in our simulations (more precisely scrambled Sobol' sequences;
see \cite{Owen1997a,Owen1997b,Owen1998} for more details on scrambling), as this makes it
possible to evaluate the numerical error through repeated runs.

Finally, while we presented SQMC in this specific case where particles are mutated according to $f^X(\bx_t|\bx_{t-1})$, the Markov transition of the considered model, it of course extends directly to situations where particles are mutated
according to some other kernel $q_t(\bx_t|\bx_{t-1})$ (assuming that the particles
are reweighted accordingly, as in standard SMC).

\section{Application: Autonomous positioning}

\subsection{Model description}

We consider the problem of autonomous positioning of a vehicle moving in a two
dimensional space. To determine its position, the vehicle estimates its speed
every $T_s>0$ seconds and measures the power of $d_y\geq 1$ radio signals. We
suppose that the radio signals are emitted from known locations
$\bm{r}_i\in\mathbb{R}^2$, $i=1,\dots,d_y$, and that the corresponding
attenuation coefficients $\alpha_i$ are known as well. This positioning problem
admits the following state space representation (see \cite{Gustafsson2002} and
\cite{Miguez2004})
\begin{equation}\label{eq:Model}
\begin{cases}
y_{ti}=10\log_{10}\Big(\frac{P_{i0}}{\|r_i-\bx_t\|^{\alpha_i}}\Big)+\nu_{it},
&t\geq 0\\ \bx_t=\bx_{t-1}+T_s\bm{v}_t+T_s\bm{\epsilon}_t,& t\geq 1\\
\bx_0\sim\mathcal{N}_2(\bm{0},\bm{I}_2)
\end{cases}
\end{equation} where $i\in 1:d_y$, $\bx_t\in\mathbb{R}^2$ is the position of the
vehicle at time $t$, $\bm{v}_t$ is a measure of its speed, which is assumed to
be constant over successive time intervals of $T_s$ seconds, $\bm{\epsilon}_t$
and $\bm{\nu}_y=(\nu_{1t},\dots,\nu_{d_yt})$ represent measurement errors while
$y_{it}$ is the power received at time $t$ by emitter $i$.  In the sequel,
$P_{0i}$ is the initial signal from emitter $i$ and, following
\cite{Miguez2004}, we suppose that all the error terms are independent and
distributed according to a Laplace distribution with parameter 0.5.

\subsection{Simulation set-up}
\label{sec:simulation-set-up}
\begin{figure} \centering
\includegraphics[scale=0.35]{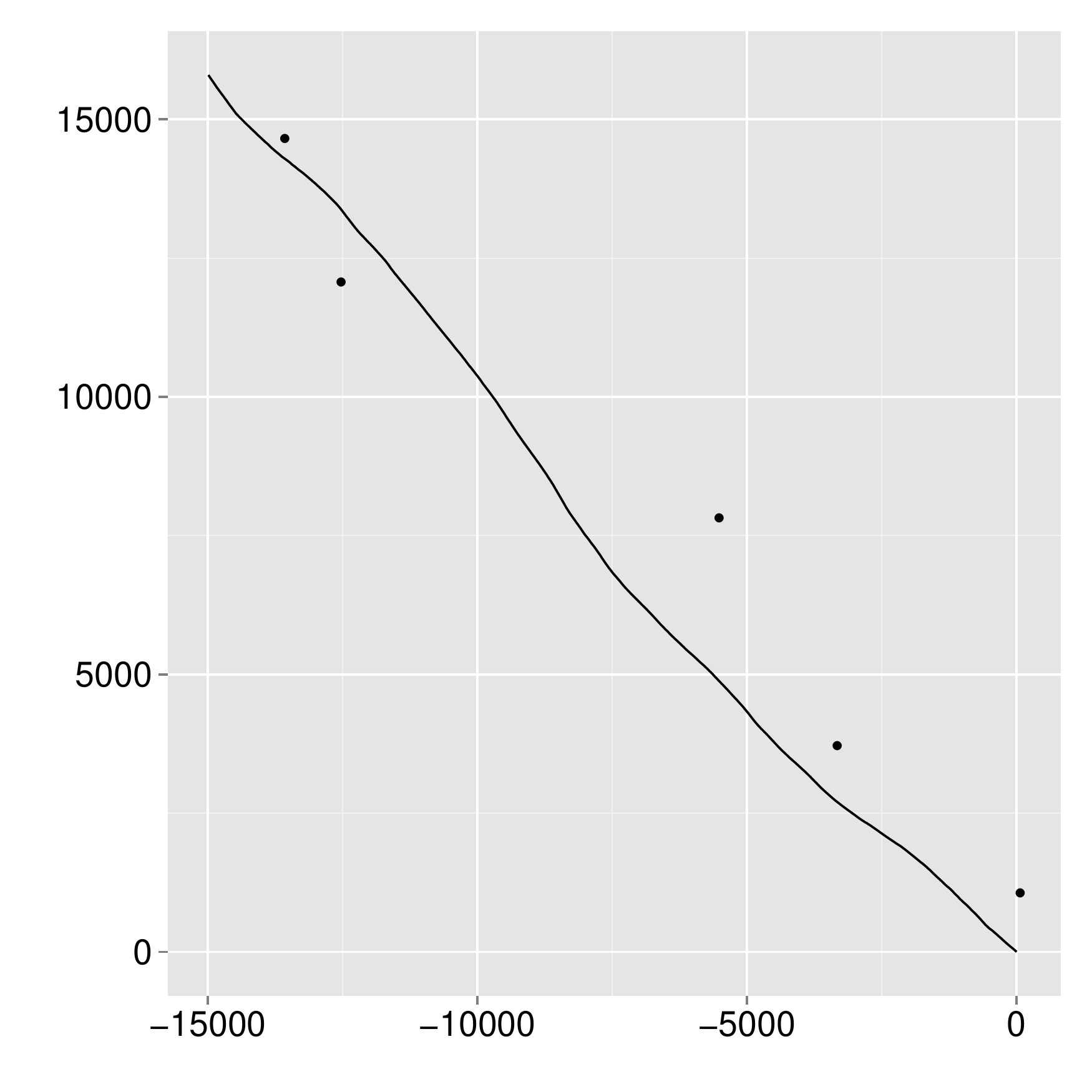}
\caption{Trajectory of a vehicle evolving for 15 minutes and starting at a
location close to $(0,0)$. The dots show the locations of the 5
emitters.\label{fig:Trajectory}}
\end{figure}

To compare the performance of SQMC and SMC for this tracking problem
we simulate the trajectory of a vehicle evolving during 15 minutes
according to \eqref{eq:Model}. We assume that the sample period is
$T_s=1$ second, that $d_y=5$ (5 emitters) and that $\alpha_i=0.95$ for
all $i=1,\dots,d_y$. The resulting trajectory and the locations of the
emitters are shown in Figure \ref{fig:Trajectory}.

The SMC algorithm is implemented using systematic resampling
\cite{CarClifFearn}, which is usually recognised as being the most
efficient resampling strategy.

SQMC is implemented using nested scrambled Sobol' sequences for the
point sets $\bu_t^{1:N}$. As described above, we need to use a mapping
$\psi_t$ to map the particles generated at iteration $t$ of SQMC into
the unit square before performing the Hilbert sort. Following
\cite{SQMC}, we chose for $\psi_t$ a component-wise (re-scaled)
logistic transform; that is, $
\psi_t(\bx)=(\psi_{t1}(x_1),\psi_{t2}(x_2)) $ with
$$
\psi_{ti}(x_i)=\left[1+\exp\left(-\frac{x_{i}-\underline{x}_{ti}}{\bar{x}_{ti}-\underline{x}_{ti}}\right)\right]^{-1},\quad
i=1,2.
$$
and where the time varying constants $\bar{x}_{ti}$ and $\underline{x}_{ti}$ are
used to solve numerical problems due to high values of $|x_i|$. More precisely,
these constants should be chosen such that, with high probability, $x_{ti}\in
[\underline{x}_{ti},\bar{x}_{ti}]$. 
To this aims, note that
\begin{align*} \var(x_{ti})&=\var(\bx_{0i})+t\, T_s^2\,\var(\epsilon_{1i}).
\end{align*} 
and thus, a reasonable choice for $\underline{x}_{ti}$ and
$\bar{x}_{ti}$ is
\begin{align*} &\underline{x}_{ti},\,\bar{x}_{ti}=\sum_{s=0}^t\bm{v}_s\pm
2\sqrt{\var(\bx_{0i})+t\, T_s^2\, \var(\epsilon_{1i})}.
\end{align*}
Simulation results are presented for $N\in\{2^8,\dots,2^{16}\}$, where $2$ is the base of the Sobol' sequence. Taking a power of 2 for the number of simulations is the standard approach in QMC integration based on Sobol' sequence because both good theoretical  and empirical results are obtained for this choice of $N$. However, this restriction is non necessary for QMC to outperform Monte Carlo methods and little gain may be expected in the context of SQMC, see \cite{Gerber:QMCarbitrarysize} for more details on this point.

\subsection{Results}

\begin{figure} \centering
\includegraphics[scale=0.35]{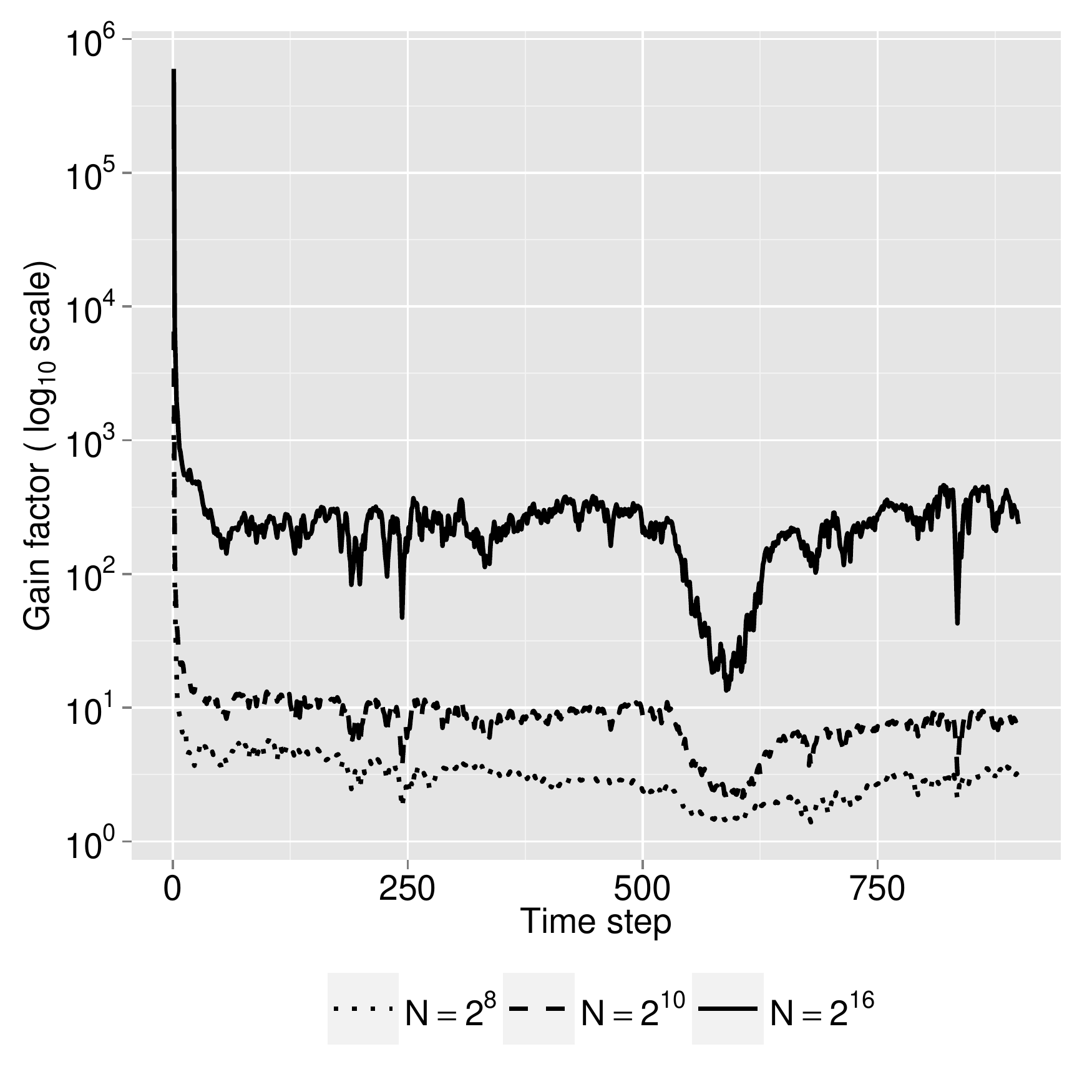}
\caption{Filtering of the state-space model \eqref{eq:Model}: The plot gives the
gain factor, defined as the MSE(SMC) over MSE(SQMC), as a function of $t$ for
the estimation of $\E[x_{1t}|\by_{0:t})]$. The results are obtained from 100
independent runs of SMC and SQMC.\label{fig:GainFactor}}
\end{figure}

In Figure \ref{fig:GainFactor} we compare the mean square error (MSE) of the
filtering expectation estimate obtained from SQMC and SMC, as a function of $t$, for
$N\in\{2^8,2^{10},2^{16}\}$.  To save space, only the results for the first component
of $\bx_t$ are presented; the results for the second component are essentially the
same. One observes that the performance gain of SQMC (relative to standard particle
filtering) increases quickly with $N$.

We now study the amount of CPU time required to have a ``reasonable'' Monte Carlo
error using both SMC and SQMC. Letting $\hat{\bx}_t$ be an estimate of the filtering
expectation $\E[\bx_t|\by_{0:t}]$, we consider the Monte Carlo error to be reasonable
if it is small compared to the posterior variance, that is, if
$\mathrm{MSE}(\hat{x}_{it})\leq \delta^2\var(x_{it}|\by_{0:t})$ for $i=1,2$ and where
we set $\delta=0.01$.

\begin{figure} \centering
\includegraphics[scale=0.35]{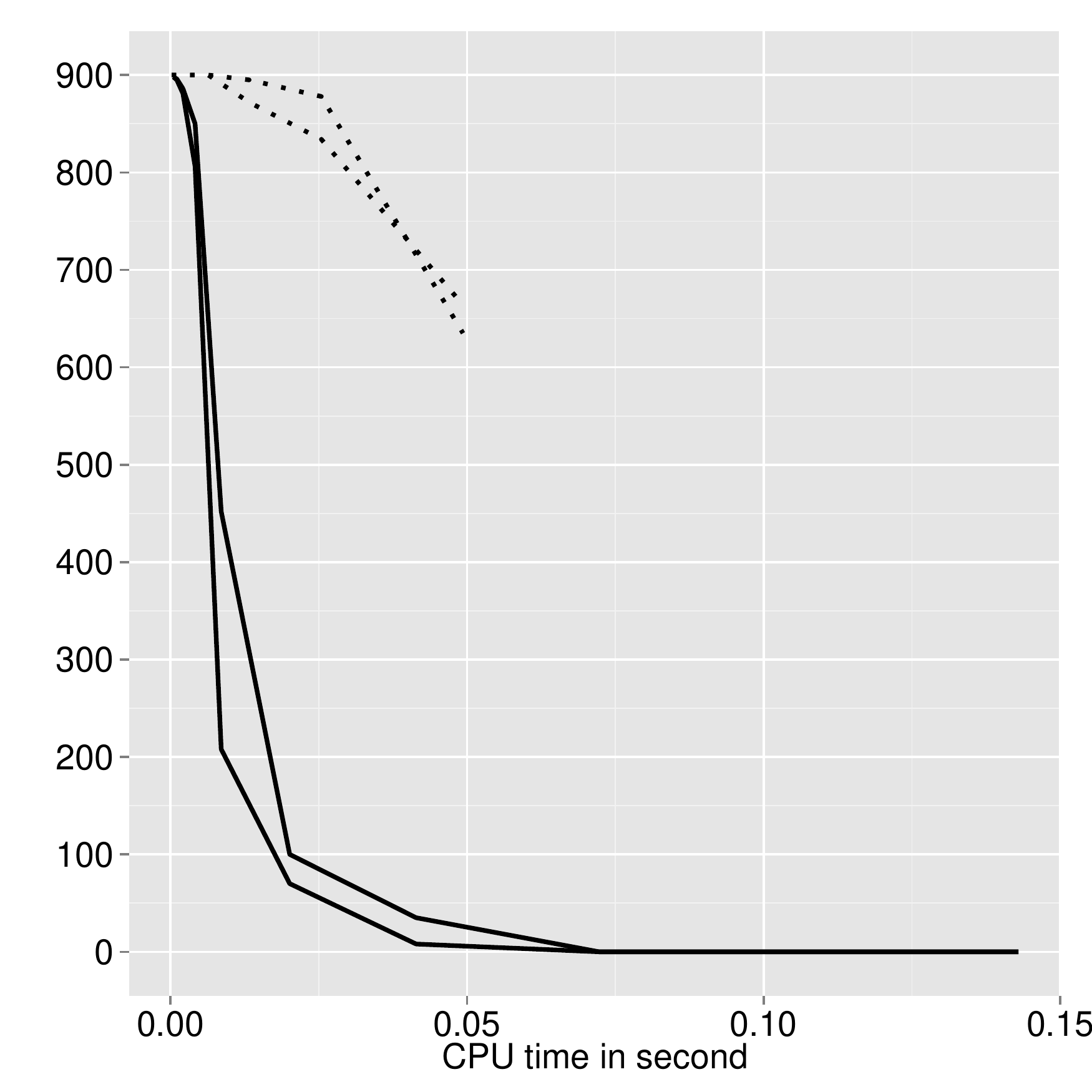}
\caption{Filtering of the state-space model \eqref{eq:Model}: The plot gives the
number of time steps $t\in\{0,\dots,899\}$ such that $\mathrm{MSE}(\hat{x}_{ti})\geq
0.01^2\var(x_{ti}|\by_{0:t})$ as a function of the CPU budget (average CPU time per
iteration), where $\hat{x}_{it}$ is either the SQMC (solid lines) or the SMC
(dashed lines) estimate of $\E[x_{ti}|\by_{0:t}]$, $i=1,2$. The results are
obtained from 100 independent runs of SMC and SQMC.\label{fig:DeltaCPU}}
\end{figure}

Figure \ref{fig:DeltaCPU} shows the number of time steps $t\in\{0,\dots,899\}$ for
which this criterion is not met, as a function of the CPU budget (i.e.  CPU time per
iteration).  To increase the CPU budget, we simply increase $N$.  We observe that
much better results are achieved using SQMC. Indeed, when the CPU budget is 0.05s 
per iteration, the SMC error is too large for more than 600 time steps, while a CPU
budget of 0.07s is enough to estimate both coordinates of $\bx_t$ for all
iterations with SQMC.

\section{Conclusion}

In this paper we have  illustrated the potential of sequential quasi-Monte Carlo for real time  signal processing processing problems with a non-linear and non-Gaussian state-space model for autonomous positioning. Compared to Monte Carlo particle filtering, dramatic variance reductions are observed when SQMC is used, both as a function of the number of particles and of CPU time. In real time application, the running time of the filtering algorithm is a crucial element and, concerning this point, we believe that significant improvement can be achieved for SQMC, notably concerning the Hilbert sort step. For instance, the computations of the Hilbert indices involve only bits operations and therefore GPU computing may allow for dramatic cost reductions.

\bibliographystyle{IEEEbib} \bibliography{complete}

\begin{thebibliography}{10}

\bibitem{SQMC}
M.~Gerber and N.~Chopin,
\newblock ``Sequential quasi-{M}onte {C}arlo,''
\newblock {\em J. R. Statist. Soc. B (to appear)}, 2015.

\bibitem{DouFreiGor}
A.~Doucet, N.~de~Freitas, and N.~J. Gordon,
\newblock {\em Sequential {M}onte {C}arlo Methods in Practice},
\newblock Springer-Verlag, New York, 2001.

\bibitem{CapMouRyd}
O.~Capp\'e, E.~Moulines, and T.~Ryd\'en,
\newblock {\em Inference in Hidden {M}arkov Models},
\newblock Springer-Verlag, New York, 2005.

\bibitem{Djuric2004}
P.~M. Djuric, S.~Godsill, and A.~Doucet,
\newblock ``Special issue on particle filtering in signal processing,''
\newblock {\em EURASIP Journal of Applied Signal Processing}, 2004.

\bibitem{Miguez2004}
J.~M{\'\i}guez, M.~F. Bugallo, and P.~M. Djuri{\'c},
\newblock ``A new class of particle filters for random dynamic systems with
  unknown statistics,''
\newblock {\em EURASIP Journal on Applied Signal Processing}, vol. 2004, pp.
  2278--2294, 2004.

\bibitem{Lemieux:MCandQMCSampling}
Christiane Lemieux,
\newblock {\em {Monte Carlo and Quasi-Monte Carlo Sampling (Springer Series in
  Statistics)}},
\newblock Springer, February 2009.

\bibitem{dick2010digital}
J.~Dick and F.~Pillichshammer,
\newblock {\em Digital nets and sequences: discrepancy theory and quasi-Monte
  Carlo integration},
\newblock Cambridge University Press, 2010.

\bibitem{Owen1995}
A.~B. Owen,
\newblock ``Randomly permuted $(t, m, s)$-nets and (t, s)-sequences,''
\newblock in {\em Monte Carlo and Quasi-Monte Carlo Methods in Scientific
  Computing. Lecture Notes in Statististics}, vol. 106, pp. 299--317. Springer,
  New York, 1995.

\bibitem{Owen1997b}
A.~B. Owen,
\newblock ``Scramble net variance for integrals of smooth functions,''
\newblock {\em Ann. Stat.}, vol. 25, no. 4, pp. 1541--1562, 1997.

\bibitem{Hamilton2008b}
Chris~H. Hamilton and Andrew Rau-Chaplin,
\newblock ``Compact {H}ilbert indices: Space-filling curves for domains with
  unequal side lengths,''
\newblock {\em Inf. Process. Lett.}, vol. 105, no. 5, pp. 155--163, 2008.

\bibitem{Owen1997a}
A.~B. Owen,
\newblock ``Monte {C}arlo variance of scrambled net quadrature,''
\newblock {\em SIAM Journal on Numerical Analysis}, vol. 34, no. 5, pp.
  1884--1910, 1997.

\bibitem{Owen1998}
A.~B. Owen,
\newblock ``Scrambling {S}obol' and {N}iederreiter-{X}ing points,''
\newblock {\em Journal of complexity}, vol. 14, no. 4, pp. 466--489, 1998.

\bibitem{Gustafsson2002}
F.~Gustafsson, F.~Gunnarsson, N.~Bergman, U.~Forssell, J.~Jansson, R.~Karlsson,
  and P.-J. Nordlund,
\newblock ``Particle filters for positioning, navigation, and tracking,''
\newblock {\em Signal Processing, IEEE Transactions on}, vol. 50, no. 2, pp.
  425--437, 2002.

\bibitem{CarClifFearn}
J.~Carpenter, P.~Clifford, and P.~Fearnhead,
\newblock ``Improved particle filter for nonlinear problems,''
\newblock {\em IEE Proc. Radar, Sonar Navigation}, vol. 146, no. 1, pp. 2--7,
  1999.

\bibitem{Gerber:QMCarbitrarysize}
M.~Gerber,
\newblock ``On integration methods based on scrambled nets of arbitrary size,''
\newblock {\em arXiv:1408.2773}, 2014.

\end{thebibliography}

\end{document}